\documentclass{PoS}

\title{Highlights from the VERITAS AGN Observation Program}

\ShortTitle{Highlights from the VERITAS AGN Observation Program}

\author{\speaker{Wystan Benbow} for the VERITAS Collaboration\thanks{https://veritas.sao.arizona.edu/}\\
        Center for Astrophysics | Harvard \& Smithsonian, 60 Garden St, Cambridge, MA, 02138, USA\\
        E-mail: \email{wbenbow@cfa.harvard.edu}}

\abstract{VERITAS is one of the world's most sensitive detectors of
  astrophysical VHE (E $>$ 100 GeV) gamma rays.  This array of four 12-m
  imaging atmospheric-Cherenkov telescopes has operated for $\sim$12 years,
  and  $\sim$6,000 hours of observations have been targeted on active
  galactic nuclei (AGN).  Approximately 300 AGN have been observed
  with VERITAS, and 39 are detected.  Most of these detections are
  accompanied by contemporaneous, broadband observations, which enable
  detailed studies of the underlying jet-powered processes. Recent
  highlights from the VERITAS AGN observation program and scientific
  results are presented.}

\FullConference{36th International Cosmic Ray Conference -ICRC2019-\\
		July 24th - August 1st, 2019\\
		Madison, WI, U.S.A.}

\begin{document}

\section{Introduction}

AGN comprise about one-third of the VHE sky catalog and they
are the most numerous class of identified VHE $\gamma$-ray emitter.
There are 78 VHE-detected AGN as of  {\it ICRC2019} \cite{TEVCAT}, and $\sim$75\%
of these objects have Northern declinations ($\delta > 0^{\circ}$).
Accordingly they are prime targets for Northern VHE observatories
such as VERITAS \cite{VERITAS_spec}.  The observed emission from VHE
AGN is dominantly non-thermal and is characterized by a double-humped spectral energy distribution (SED)
that spans the entire broadband spectrum.  This radiation is highly
variable at all wavelengths, and correspondingly most modern efforts to
probe their underlying processes require contemporaneous
multi-wavelength (MWL) observations.  The VHE $\gamma$-ray emission in AGN is
generally produced by accretion-powered jets in a compact region near their central, supermassive black hole.  
Decades after this VHE emission was first discovered, its origin
remains debated, although processes involving leptonic particles (e.g. synchrotron self-Compton models)
are typically favored over ones focusing on contributions from hadronic particles.

Blazars, objects whose jets are pointed close to the line of sight
towards Earth, form the dominant class ($\sim$95\%) of VHE AGN. Four 
nearby ($z < 0.022$) FR-I radio galaxies are also detected in the VHE
band.  However, these objects are not strongly misaligned and
it is debated whether another two relatively nearby VHE AGN are radio galaxies or
blazars. Among the 72 VHE AGN that are certainly blazars, a majority
(62) are BL Lac objects; the rest (7) are either Flat Spectrum
Radio Quasars (FSRQs) or have uncertain blazar sub-classification (3).
The detected BL Lac objects are further sub-classified based on the
location of their lower-energy SED peak, and 51 are
high-frequency-peaked (HBLs), 8 are intermediate-frequency-peaked
(IBLs) and 2 are low-frequency-peaked (LBLs).  Unfortunately the
defining characteristic of BL Lac objects (i.e. the absence of
features in their optical spectra) implies difficulties
in measuring their redshift.  The VHE blazar catalog currently covers
a redshift range from $z = 0.030$ to $z = 0.954$, but at least
$\sim$20\% of the objects have uncertain redshift. Only $\sim$20\% of
the catalog has $z >0.3$ and more than 50\% have $z < 0.2$.  While energetics requirements certainly
contribute to general proximity of the VHE AGN catalog, the attenuation of
VHE photons in a distance- and energy-dependent manner by the
extragalactic background light is also a major effect \cite{Elisa_ICRC}.

Empirically there are two qualities that typically characterize VHE
AGN and this drives the design of the VERITAS AGN Program (radio galaxies and
blazars). First and foremost, their observed VHE flux is almost always variable.
About one-third of VHE AGN are only detected during short-duration
flares, and major AGN outbursts are the defining characteristic of the
VHE field for many  (see, e.g., \cite{PKS2155_flare}).  However, it is
also important to note that particularly notable episodes of 
rapid (minute-scale), large-scale (factor of 100) flux
variations are very rare and most VHE flux variations are of the order
of a factor of 2-3.  In general the time-scales observed for these
milder variations depends on the average VHE flux.  They can be as long as an observing season,
but only brighter objects show such variations on shorter time
scales and this is very likely an instrument-sensitivity effect.  The
VERITAS AGN
Program therefore attempts to identify and follow-up on VHE AGN
flares, noting that variations of even a factor 2$-$3 can be very
interesting for some targets.
The other important empirical quality is that the observed photon spectra of
VHE AGN are often soft ($\Gamma_{obs} \sim 3 - 5$), and very few VHE
AGN are detected above 1 TeV. While this can be related to the
higher-energy SED peak location, it is also in part due to EBL-absorption
effects. In most cases, the harder the VHE spectrum is, and the
higher energy to which it reaches, the more interesting the target
becomes scientifically.  Therefore the VERITAS AGN Program also
focuses on hard spectrum VHE blazars and generating statistics above
above 1 TeV.

In general, the goal of the VERITAS program is to make precision measurements 
of VHE AGN spectra and their variability patterns.  It necessarily leverages
contemporaneous MWL observations from both ground-
and space-based facilities, in particular the FLWO 48''
optical telescope and Swift XRT/UVOT.  Since transitioning from an
emphasis on expanding the VHE AGN catalog in $\sim$2010, 
the program has focused on long-term studies of known VHE AGN
population in a manner that emphasizes the regular search for, and intense
observation of, major flaring episodes.  From 2013-2018, the program
sampled each Northern VHE AGN ($\sim$56), but in 2018 it was
streamlined again to more heavily focus on more intense studies of a
few ($\sim$22) targets. Independent of any successful
flare identification, the regular sampling of each
VHE AGN built high-statistics data sets to enable fully-constrained modeling of each
VHE AGN's SED (see, e.g., \cite{1ES0229_paper}), and the ongoing
sampling continues to improve these.
The various long-term MWL light curves should also allow
for flux and spectral correlation studies that may indicate
commonalities in the origin of each AGN's emission. In generating a
large sample of precision VHE AGN spectra, the VERITAS AGN program
is also useful for generating a variety of cosmological measurements
such as constraints on the the density of the EBL \cite{Elisa_ICRC}
and  the strength of the intergalactic magnetic field (IGMF) \cite{VERITAS_IGMF}.

\begin{figure*}[t]
   \centerline{ {\includegraphics[width=3.5in]{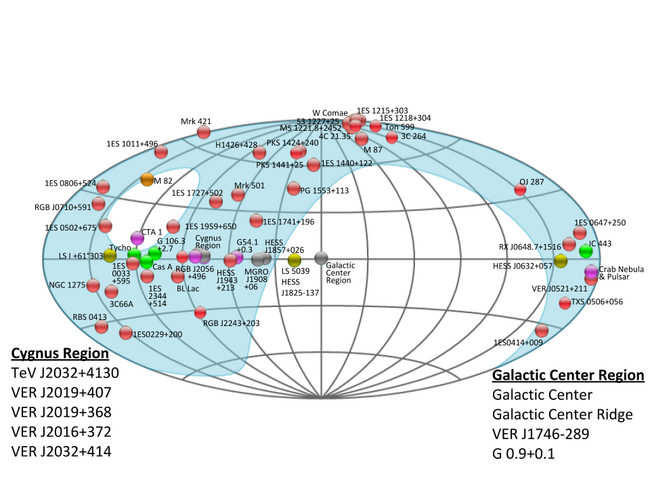} } }
\vspace{-0.3cm}
   \caption{{\footnotesize The VERITAS sky catalog in Galactic
       coordinates.  The blue region is the sky visible to VERITAS at
       zenith angles $<$35$^{\circ}$. Different astrophysical classes are shown with
       different colored markers.  The red circles are AGN. }}
   \label{VERITAS_catalog}
\vspace{-0.3cm}
 \end{figure*}

\begin{table}[t]
\begin{center}
\begin{tabular}{c | c | c | c |c}
\hline
{\footnotesize AGN} & {\footnotesize $z$} &  {\footnotesize Type} & {\footnotesize log$_{10}(\nu_{\rm synch})$ [Hz]} & {\footnotesize TeV FoM}\\
\hline
\vspace{-0.13cm}
{\footnotesize M\,87} & {\footnotesize 0.004} & {\footnotesize FR\,I} & {\footnotesize $--$} & {\footnotesize $--$}\\\
\vspace{-0.13cm}
{\footnotesize NGC\,1275} & {\footnotesize 0.018} & {\footnotesize FR\,I} & {\footnotesize $--$} & {\footnotesize $--$}\\\
\vspace{-0.13cm}
{\footnotesize 3C\,264$^{\dagger}$} & {\footnotesize 0.022} & {\footnotesize FR\,I} & {\footnotesize $--$} & {\footnotesize $--$}\\\
\vspace{-0.13cm}
{\footnotesize Mrk\,421} & {\footnotesize 0.030} & {\footnotesize HBL} & {\footnotesize 16.3} & {\footnotesize 39.8}\\\
\vspace{-0.13cm}
{\footnotesize Mrk\,501} & {\footnotesize 0.034} & {\footnotesize HBL} & {\footnotesize 17.9} & {\footnotesize 12.6}\\\
\vspace{-0.13cm}
{\footnotesize 1ES\,2344+514} & {\footnotesize 0.044} & {\footnotesize HBL} & {\footnotesize 17.7} & {\footnotesize 1.58}\\\
\vspace{-0.13cm}
{\footnotesize 1ES\,1959+650} & {\footnotesize 0.047} & {\footnotesize  HBL} & {\footnotesize 17.0} & {\footnotesize 20.0}\\
\vspace{-0.13cm}
{\footnotesize 1ES\,1727+502} & {\footnotesize 0.055} & {\footnotesize  HBL} & {\footnotesize 17.0} & {\footnotesize 3.16}\\
\vspace{-0.13cm}
{\footnotesize BL\,Lac} & {\footnotesize 0.069} & {\footnotesize  IBL} & {\footnotesize 14.3*} & {\footnotesize $--$}\\
\vspace{-0.13cm}
{\footnotesize 1ES\,1741+196} & {\footnotesize 0.084} & {\footnotesize  HBL} & {\footnotesize 17.8} & {\footnotesize 1.00}\\
\vspace{-0.13cm}
{\footnotesize W\,Comae$^{\dagger}$} & {\footnotesize 0.102} & {\footnotesize IBL} & {\footnotesize 14.8*} & {\footnotesize $--$}\\
\vspace{-0.13cm}
{\footnotesize RGB\,J0521.8+2112$^{\dagger}$} & {\footnotesize 0.108} & {\footnotesize IBL} & {\footnotesize 15.1} & {\footnotesize 5.01}\\
\vspace{-0.13cm}
{\footnotesize RGB\,J0710+591$^{\dagger}$} & {\footnotesize 0.125} & {\footnotesize HBL} & {\footnotesize 18.1} & {\footnotesize 3.98}\\
\vspace{-0.13cm}
{\footnotesize H\,1426+428} & {\footnotesize 0.129} & {\footnotesize HBL} & {\footnotesize 18.1} & {\footnotesize 3.98}\\
\vspace{-0.13cm}
{\footnotesize B2\,1215+30} & {\footnotesize 0.131} & {\footnotesize HBL} & {\footnotesize 15.1} & {\footnotesize 2.51}\\
\vspace{-0.13cm}
{\footnotesize S3\,1227+25$^{\dagger}$} & {\footnotesize 0.135} & {\footnotesize IBL} & {\footnotesize 15.0} & {\footnotesize 3.98}\\
\vspace{-0.13cm}
{\footnotesize 1ES\,0806+524$^{\dagger}$} & {\footnotesize 0.138} & {\footnotesize HBL} & {\footnotesize 15.9} & {\footnotesize 3.16}\\
\vspace{-0.13cm}
{\footnotesize 1ES\,0229+200} & {\footnotesize 0.140} & {\footnotesize HBL} & {\footnotesize 18.5} & {\footnotesize 2.00}\\
\vspace{-0.13cm}
{\footnotesize 1ES\,1440+122$^{\dagger}$} & {\footnotesize 0.163} & {\footnotesize HBL} & {\footnotesize 17.2} & {\footnotesize 1.26}\\
\vspace{-0.13cm}
{\footnotesize RX\,J0648.7+1516$^{\dagger}$} & {\footnotesize 0.179} & {\footnotesize HBL} & {\footnotesize 16.6} & {\footnotesize 1.58}\\
\vspace{-0.13cm}
{\footnotesize 1ES\,1218+304} & {\footnotesize 0.182} & {\footnotesize HBL} & {\footnotesize 16.8} & {\footnotesize 3.16}\\
\vspace{-0.13cm}
{\footnotesize RBS\,0413$^{\dagger}$} & {\footnotesize 0.190} & {\footnotesize HBL} & {\footnotesize 17.3} & {\footnotesize 1.00}\\
\vspace{-0.13cm}
{\footnotesize 1ES\,1011+496} & {\footnotesize 0.212} & {\footnotesize HBL} & {\footnotesize 16.2} & {\footnotesize 3.98}\\
\vspace{-0.13cm}
{\footnotesize MS\,1221.8+2452} & {\footnotesize 0.218} & {\footnotesize HBL} & {\footnotesize 16.1} & {\footnotesize 1.26}\\
\vspace{-0.13cm}
{\footnotesize 1ES\,0414+009} & {\footnotesize 0.287} & {\footnotesize  HBL} & {\footnotesize 17.9} & {\footnotesize 3.16}\\
\vspace{-0.13cm}
{\footnotesize OJ\,287$^{\dagger}$} & {\footnotesize  0.306} & {\footnotesize BL\,Lac} & {\footnotesize 13.9*} & {\footnotesize $--$}\\
\vspace{-0.13cm}
{\footnotesize TXS\,0506+056} & {\footnotesize  0.337} & {\footnotesize Blazar} & {\footnotesize 15.3*} & {\footnotesize $--$}\\
\vspace{-0.13cm}
{\footnotesize 1ES\,0502+675$^{\dagger}$} & {\footnotesize 0.341} & {\footnotesize HBL} & {\footnotesize 17.9} & {\footnotesize 3.98}\\
\vspace{-0.13cm}
{\footnotesize PKS\,1222+216} & {\footnotesize 0.432} & {\footnotesize FSRQ} & {\footnotesize $--$} & {\footnotesize $--$}\\
\vspace{-0.13cm}
{\footnotesize 1ES\,0033+595} & {\footnotesize 0.467} & {\footnotesize HBL} & {\footnotesize 18.9*} & {\footnotesize $--$}\\
\vspace{-0.13cm}
{\footnotesize PKS\,1424+240$^{\dagger}$} & {\footnotesize 0.604} & {\footnotesize HBL} & {\footnotesize 15.0}  & {\footnotesize 7.94}\\
\vspace{-0.13cm}
{\footnotesize Ton 599$^{\dagger}$} & {\footnotesize 0.7247} & {\footnotesize FSRQ} & {\footnotesize $--$} & {\footnotesize $--$}\\
{\footnotesize PKS\,1441+25} & {\footnotesize 0.939} & {\footnotesize FSRQ} & {\footnotesize $--$} & {\footnotesize $--$}\\
\hline
\hline
\vspace{-0.13cm}
{\footnotesize 3C\,66A$^{\dagger}$} & {\footnotesize  $0.33 < z <  0.41$} & {\footnotesize IBL} & {\footnotesize 15.6*} & {\footnotesize $--$}\\
\vspace{-0.13cm}
{\footnotesize 1ES\,0647+250} & {\footnotesize ?} & {\footnotesize HBL} & {\footnotesize 16.8} & {\footnotesize 3.16}\\
\vspace{-0.13cm}
{\footnotesize PG\,1553+113} & {\footnotesize $0.43 < z < 0.50$} & {\footnotesize HBL} & {\footnotesize 15.6} & {\footnotesize 7.94}\\
\vspace{-0.13cm}
{\footnotesize HESS\,J1943+213} & {\footnotesize ?} & {\footnotesize  HBL} & {\footnotesize 18.1} & {\footnotesize 2.00}\\
\vspace{-0.13cm}
{\footnotesize RGB\,J2056+496$^{\dagger}$} & {\footnotesize ?} & {\footnotesize Blazar} & {\footnotesize $--$} & {\footnotesize $--$}\\
{\footnotesize RGB\,J2243+203$^{\dagger}$} & {\footnotesize ?} & {\footnotesize HBL} & {\footnotesize 15.1} & {\footnotesize 1.58}\\
\hline
\end{tabular}
\vspace{-0.2cm}
\caption{{\footnotesize The 39 AGN (25 HBL, 5 IBL, 3 FSRQs, 3
    unclassified blazars, and 3 radio galaxies)
    detected with VERITAS. This catalog has grown by 3, 5, 7, 13, and
    22 AGN since each of the previous 5 ICRCs, respectively.  The
    16 blazars discovered at VHE by VERITAS are marked with a dagger. The
    classifications are taken from TeVCat, and the synchrotron peak
    frequencies and TeV Figures of Merit are from \cite{2WHSP}; the
    frequencies marked with an asterisk are from \cite{Nieppola}.}}
\label{AGN_catalog}
\end{center}
\vspace{-0.9cm}
\end{table}

\section{VERITAS AGN Program}

VERITAS is the world's most sensitive observatory between $\sim$85 GeV
and $\sim$30 TeV and it is regularly used to study the Northern sky
during $\sim$10-month seasons (September $-$ July).  It began full
operations in 2007 at the F. L. Whipple Observatory in Arizona, USA  (31$^{\circ}$ 40' N, 110$^{\circ}$ 57' W,  1.3
km a.s.l.), and it achieved its present sensitivity following a series of
upgrades completed in Summer 2012.  The array of Cherenkov telescopes
can be used to detect an object with $\sim$1\% Crab Nebula flux (1\%
Crab) in less than 25 hours, and spectra can be generated above
$\sim$100 GeV. For most measurements, the systematic
errors are $\sim$0.1 on the photon index
($\Gamma$) and $\sim$20\% on the flux.

The VERITAS collaboration has acquired a total of $\sim$14,600 h of
full-scale observations, averaging $\sim$930 h
of good-weather observations each season during ``dark time'' (moon
illumination $<$30\%), and since 2012, $\sim$205 h during periods of ``bright
moonlight'' (i.e. $>$30\% illumination).
The bright-moon data has comparable sensitivity to
dark-time observations with only slightly higher threshold (e.g. 250
GeV) \cite{BrightMoon_paper}, and are useful for selected AGN targets.

Given VERITAS's Northern Hemisphere location, AGN observations are 
naturally a significant component of its data taking ($\sim$50\%).
As of July 2019, these data comprise a total of $\sim$6,100 h ($\sim$425 h per
year) of good-weather dark time and $\sim$1,000 h ($\sim$170 h per year) of
good-weather, bright-moon time.  The dark time is typically split  $\sim$90\% to blazars, primarily BL Lac objects, 
and $\sim$10\% to radio-galaxies. The bright-moon
time for AGN is devoted almost entirely to BL\,Lac objects, with
recent ($>$2017) observations split $\sim$45\% to candidates for new
VHE discoveries, and $\sim$55\% to known VHE blazars, particularly those with hard
VHE spectra.   This marks a shift from prior seasons where the split was
65\% / 35\%, respectively. AGN comprise 62\% of the VERITAS source
catalog (shown in Figure~\ref{VERITAS_catalog}),
and Table~\ref{AGN_catalog} lists the 39 AGN detected by VERITAS.

The VERITAS AGN program is based on regular monitoring observations of the Northern VHE catalog 
to self-identify VHE flaring episodes for immediate MWL
target-of-opportunity (ToO) follow-up.  
The monitoring program's minimum sample duration will detect $\sim$10\% Crab
flux, and the observation cadence ranges from daily to weekly.
While any monitoring observation could fortuitously catch a short-duration, bright flare (e.g. BL\,Lac in 2017
\cite{BLLAC_2017}), in general the concept is to identify
long-lasting, bright states for initiating intense campaigns.
Naturally the VERITAS monitoring observations are supplemented with
coordinated data at lower energy to assist with triggering and to ensure that long-term
contemporaneous MWL data sets exist for the monitored AGN.
From 2013-18, every VHE AGN was monitored by VERITAS, and
some AGN have been monitored continuously since $\sim$2010. Over a
long period, even the lowest level monitoring significantly increased the
data set for each object.  In 2018, the program was streamlined based
on each target's existing VHE variability profile (e.g. from the
VERITAS data), its possibilities for external triggers (e.g. from
{\it Fermi}-LAT or FACT), and its perceived scientific importance. This
streamlining from 59 to 22 targets enables much deeper, legacy 
exposures on particularly interesting objects
during what are likely to be the final years of VERITAS operations.

While the primary focus of the VERITAS AGN program is performing deep / timely measurements of
the known VHE sources, $\sim$40\% of the AGN program was devoted to the discovery and
follow-up observations of new VHE AGN between 2017-19. Most of these
observations were either ToO observations triggered by one of
our MWL partners or regular observations of targets from a list of
selected candidates.  Recently, our discovery
candidates include AGN with a weak excess ($>$3$\sigma$) in
large, archival VERITAS exposures, and those remaining in a comprehensive list of Northern objects
with only limited exposures ($<$4 h).  The comprehensive
target list was generated in 2016, and included all the X-ray brightest HBLs in the 2WHSP catalog (i.e. objects with a
``TeV Figure of Merit'' $>$ 1.0) ~\cite{2WHSP}, all the hardest
($\Gamma_{2FHL} < 2.8$) AGN in the {\it Fermi}-LAT 2FHL ($>$50 GeV)
catalog \cite{2FHL_Catalog}, and all targets previously selected for
VERITAS \cite{Benbow09,Benbow11}.

ToO observations are the highest priority of the VERITAS AGN  program.
Historically, these data average $\sim$25\% of the AGN dark-time yield each
season and the percentages were 26\% and only 6\% in the 2017-18 and
2018-19 seasons, respectively.  These observations
include attempts to discover and/or follow-up on new VHE
sources, and efforts to harness the potential of bright
flares in known VHE AGN.  Almost all VERITAS
FSRQ observations are taken via ToO campaigns.

\section{Recent Highlights}

{\bf TXS\,0506+056} is currently one of the most important objects in
multi-messenger astronomy.  Prior to the 2017-18 season, this
{\it Fermi}-LAT detected blazar was on the VERITAS discovery target list
because of its hard MeV-GeV spectrum.  It appears in both the 2FHL
($>$50 GeV) and 3FHL ($>$10 GeV) \cite{3FHL_catalog} catalogs, and it was relatively
bright (F($>$50 GeV) $\sim$ 4\% Crab) for a non-VHE-detected object. 
However, because its redshift ($z = 0.337$) was not yet measured and its synchrotron peak
location ($10^{15.3}$ Hz) was unremarkable, it was a low-priority
object, and only $\sim$1 h of VERITAS data were taken in 2016.  
On Sept. 22, 2017, IceCube detected a $\sim$290-TeV neutrino from a direction consistent
with the $\gamma$-ray blazar, leading to a global follow-up campaign \cite{0506_Science}. Prompt
observations of TXS\,0506+056 with VERITAS and MAGIC initially did not yield detections. However, follow-up
observations eventually revealed the blazar to be in an elevated $\gamma$-ray emission state.
This was initially seen by {\it Fermi}-LAT, and later MAGIC
discovered VHE emission. Upper limits were measured by several other VHE instruments,
including VERITAS, in $\sim$2 weeks of follow-up observations, likely due to short term variations.
VERITAS carried out extended long-term follow-up observations
of the blazar ($\sim$35 h from September 23, 2017 to February 8, 2018) leading to a
detection \cite{0506_VERITAS} at a significance of 5.8 standard deviations ($\sigma$),
albeit at lower flux than detected by MAGIC. The average 
flux observed from TXS\,0506+056 was F($>$110 GeV) = ($8.9 \pm 1.6)
\times 10^{-12}$ cm$^{-2}$ s$^{-1}$, or 1.6\% of the Crab Nebula flux, and the observed
photon index is $\Gamma = 4.8 \pm 1.3$.

The correlation of the high-energy neutrino with the
gamma-ray flare of TXS 0506+056 is statistically significant at the
level of 3$\sigma$, and could indicate that blazar jets accelerate cosmic rays to
at least several PeV. It also suggests that blazars are a source of
very-high-energy cosmic rays, and therefore contribute significantly
to the cosmic neutrino flux observed by IceCube.
Naturally, the observation of further correlations of high-energy neutrinos
with gamma-ray flares in blazars are required to solidify this
conclusion and this will be a major effort for VERITAS in the future.  This detection also impacts future neutrino
follow-up strategies with VERITAS, as the potential $\gamma$-ray
counterparts may be active over time periods of weeks or months, 
requiring multiple exposures.

{\bf Ton 599} is an FSRQ at redshift $z =
0.7247$.  It is a known, variable MeV-GeV $\gamma$-ray emitter, but
its {\it Fermi}-LAT spectrum ($\Gamma_{3FGL} = 2.1$) softens considerably at
higher energy.  It does not appear in the 2FHL ($>$50 GeV)
catalog, and the extrapolation of its 3FHL ($>$10 GeV) spectrum
($\Gamma_{3FHL} = 3.0$) to the VHE band (F($>$200 GeV) $\sim$ 0.2\% Crab flux)
suggests it is not normally detectable using VERITAS.  Given its large
redshift and unfavorable high-energy characteristics it was not
observed by VERITAS prior to 2017.  However in October
2017, a series of Astronomer's Telegrams reported a long-lasting,
remarkably high state in the optical, infrared, X-ray and $\gamma$-ray bands.
These reports initiated a series of VERITAS ToO observations when the source first became visible at reasonable zenith angles.
On December 15 and 16, 2017, an excess of $\gamma$-rays, corresponding to a statistical significance of $\sim$10$\sigma$, was quickly ($\sim$2 h) detected from the
direction of Ton\,599 (ATel\,$\#$11075).
The observed VHE flux is $\sim$12\% Crab and the spectrum is very soft ($\Gamma \sim 5$).
The object was detected on the same night (December 15) by MAGIC (ATel\,$\#$11061) and
was hence co-discovered at VHE by the two projects.  Ton\,599 is the
third FSRQ detected by VERITAS, and is the seventh overall at VHE, and
is one of the most distant VHE emitters known.  

{\bf 3C\,264} is an FR-I type radio galaxy at redshift $z =0.0216$.
It is considered a more distant ($\sim$6 times) analog to M\,87,
which is a well-known VHE emitter.  3C\,264 is an MeV-GeV source,
and the extrapolation of its 3FHL spectrum ($\Gamma_{3FHL} = 1.65$) to the VHE band (F($>$200 GeV) $\sim$ 1.6\% Crab flux)
suggests it should be detectable with VERITAS.  A remarkable aspect of
3C\,264 is its rapidly evolving knot structure revealed by long-term
Hubble Space Telescope observations \cite{Eileen_3C264}.  Four knots can be seen in its inner jet, of
which two are quasi-stationary and another two appear to be moving 
at superluminal speeds (7c and 1.8c, respectively). The two moving knots
are expected to interact within the next $\sim$30 years, and this
interaction could plausibly generate an outburst of VHE $\gamma$-rays.
Given several compelling motivations, 3C\,264 was observed with
VERITAS for $\sim$10 h in 2017 and a weak excess was initially observed.  A follow-up campaign was
organized for 2018, and the source was immediately detected in an
active VHE state ($\sim$1\% Crab flux) in a similar data sample ($\sim$12 h)
(ATel\,$\#$11436). This led to the initiation of a large VERITAS
campaign, and a significant MWL follow-up effort
including VLA, VLBI, Chandra, HST, Swift and various ground-based
optical facilities, with the hope of observing intra-jet phenomena. In total, $\sim$38 h of good-quality live-time were
acquired with VERITAS on 3C\,264, resulting in the detection of 9$\sigma$ excess
consistent with its SIMBAD position.
The observed VHE flux varies on monthly time scales, and was 
F($>$320 GeV) $\sim$ 0.6\% Crab flux on average in 2018, noting that
the bulk of the exposure was taken after the active state had subsided.
Similar to M\,87, the VHE spectrum for 3C\,264 is hard with $\Gamma =
2.2 \pm 0.3$.  A preliminary analysis of the MWL data
shows no evidence of a knot collision, a flare in the jet
microstructure, or even any significant brightening.  3C\,264 will
continue to be observed by VERITAS ($\sim$16 h per year) in future
seasons, and the 2019 VERITAS results are consistent with those from 2018.

{\bf 1ES\,1218+304} is one of the most useful objects to observe with VERITAS to
generate constraints on both the EBL and the IGMF due to its combination of {\it relatively} hard VHE spectrum,
bright VHE flux (typically $\sim$5\% Crab) and distant redshift ( $z =
0.182$).  Day-scale variability was observed from this HBL in 2009,
reaching a peak of $\sim$20\% Crab. Although heavily monitored with VERITAS
since 2009, it has shown no major VHE flux variations.
However, it began showing high activity in both the optical and X-ray
bands in late 2018.  VERITAS's
normal monitoring began in December 2018, and in January 2019 a
significant flare ($>$20\% Crab) was observed, triggering a ToO
campaign and high-cadence monitoring until April 2019.  During this
time, the Swift XRT also observed a historic peak in the X-ray count
rate.  The total good-quality VERITAS exposure is $\sim$12 h live time between
MJD 58461 and 58600 MJD.  An
excess with significance $\sim$19$\sigma$ is detected from the direction of 1ES\,1218+304,
corresponding an average flux of F($>$150 GeV) $\sim$ 11\% Crab,
peaking at $\sim$23\% Crab.  The VHE spectrum observed during
the campaign ($\Gamma \sim 3.25$) and on the flare night  ($\Gamma
\sim 3.09$) are consistent with prior observations.  More details on
the VERITAS and MWL observations of this flare can
be found in these proceedings \cite{Ste_1218}. 

\section{Conclusion}

As of July 2019, the VERITAS collaboration have acquired more 
than 14,600 hours of scientific observations, of which $\sim$12,600 h
are in good weather, and more than 6,000 of these good-weather hours are targeted on AGN.  
Since  {\it ICRC2017} the array was used to acquire $\sim$1120 h and
$\sim$820 h of good-weather observations in 2016-17 and 2017-18, respectively.
The most recent season was particularly affected by poor weather
conditions in southern Arizona.  Regardless, the good weather AGN
yields were strong in each season: $\sim$560 h and  $\sim$450 h, respectively.

During the past two seasons, the VERITAS AGN program focused heavily on regular VHE and MWL monitoring
of known Northern VHE AGN, and particularly emphasized immediate and intense ToO follow-up of interesting
flaring events. We also maintained a program focused on the discovery
of new VHE AGN, and $\sim$40\% of our most recent observations had a
discovery focus.  The most recent AGN observations resulted in the VHE
discovery of 1 radio galaxy, the co-discovery of a VHE FSRQ, the VHE detection of a BL\,Lac
object during MWL flaring possibly associated with a high-energy
neutrino emission, and the detection of
several VHE flares.  The VERITAS
catalog now includes 31 BL Lac objects, 3 FSRQs, 2 unclassified
blazars and 3 FR I radio galaxies. 

The VERITAS collaboration plans to operate the telescope array through
at least 2022 and has secured the necessary site-operations funding to
do so. Although the array is now $\sim$12 years old, it continues to run well
with the past two seasons each among historical bests for various technical
performance benchmarks (e.g. fewest hours lost to technical issues,
highest percentages of data with all telescopes operational). As the 
VERITAS collaboration's long-term science plan continues to prioritize 
AGN observations and the array continues to perform exceptionally
well, we expect our long tradition of producing exciting results to continue.

\vspace{0.1cm}

\footnotesize{This research is supported by grants from the U.S. Department of Energy Office of Science, the U.S. National Science Foundation and the Smithsonian Institution, and by NSERC in Canada. This research used resources provided by the Open Science Grid, which is supported by the National Science Foundation and the U.S. Department of Energy's Office of Science, and resources of the National Energy Research Scientific Computing Center (NERSC), a U.S. Department of Energy Office of Science User Facility operated under Contract No. DE-AC02-05CH11231. We acknowledge the excellent work of the technical support staff at the Fred Lawrence Whipple Observatory and at the collaborating institutions in the construction and operation of the instrument.}

\end{document}